# Altered temporal scaling of semantic similarity fluctuations in psychosis

**Author Names and Affiliations:** Paola Moreno Ancalmo[1], Irem Topal[1], Emre Bora[2,3,4], Rieke Roxanne Mülfarth[5,6], Svenja Seuffert[5,6], Tilo Kircher[5,6], Frederike Stein[5,6], Philipp Homan[7,8], & Wolfram Hinzen[1,9]

[1] Department of Translation & Language Sciences, Universitat Pompeu Fabra, Carrer Roc Boronat, 138,5 Barcelona, 08018, Spain

2 Department of Neurosciences, Health Sciences Institute, Dokuz Eylul University, Izmir, Turkey

3 Department of Psychiatry, Dokuz Eylul University, Izmir, Turkey

4 Department of Psychiatry, Melbourne Neuropsychiatry Centre, University of Melbourne and Melbourne Health, Carlton South, 3053, VIC, Australia

5 Department of Psychiatry and Psychotherapy, University of Marburg, Marburg, Germany

6 Center for Mind, Brain and Behavior (CMBB), Hans-Meerwein-Str. 6, 35032, Marburg, Germany

7 Department of Adult Psychiatry and Psychotherapy, University of Zurich, Zurich, Switzerland

8 Neuroscience Center Zürich, University of Zürich and ETH Zürich, Zürich, Switzerland

9 Institució Catalana de Recerca i Estudis Avançats (ICREA), Barcelona, Spain

**Funding statement.** Funded by the European Union (GA Nº 101118756 - DELTA-LANG). Views and opinions expressed are however those of the author(s) only and do not necessarily reflect those of the European Union or the Agency. Neither the European Union nor the granting authority can be held responsible for them. This work was also supported by the German Research Foundation (DFG) through grant STE3301/1-1 (project number 527712970) and by the Von Behring-Röntgen Society (project number 72_0013), both awarded to Frederike Stein. Tilo Kircher is funded by the Wellcome Trust funded DIALOG consortium (Understanding Disorganisation: A Language-Focused Global Initiative in Psychosis; co-PI: Tilo Kircher; Grant Number 314138/Z/24/Z) and the DFG-funded research unit FOR2107 (project number 240413749, KI588/14-1, KI588/14-2,

KI588/20-1, KI588/22-1). Tilo Kircher and Frederike Stein also received support from the DFG Collaborative Research Centre/Transregio 393 (CRC/TRR 393, project number 521379614).

**Abstract**. Scale-free temporal organization, in which fluctuations show similar statistical structures across timescales, is a feature of many complex systems considered to support optimal functioning. Alterations of such scaling have been reported in several pathologies, but whether comparable alterations characterize the temporal organization of language remains largely unexplored. We address this question in the domain of psychosis, where language anomalies are pervasively documented. Specifically, we apply detrended fluctuation analysis (DFA) to quantify temporal scaling in Qwen3-derived continuous cosine similarity time series, capturing trajectories through semantic space computed from transcripts of picture descriptions across three independent datasets. DFA scaling exponents ($\alpha$) were extracted to characterize the strength of long-range temporal correlations. Across all datasets, patients exhibited significantly elevated $\alpha$ relative to controls, indicating increased temporal persistence in semantic fluctuations. This analysis opens a window into the multi-timescale organization of meaning as it unfolds in discourse and reveals a signature of altered temporal scaling. This alteration parallels known departures from critical scaling in brain dynamics in psychosis, suggesting the exploration of possible links between these two.

## 1. Introduction

Language production is a fundamentally temporal process. Meaning unfolds over time and over multiple timescales, from single morphemes to phrases to sentences to discourse and conversation. These timescales are not independent from each other but hierarchically organized, inviting the study of potential self-repeating properties across scales. Patterns of self-similarity are characterized by scaling laws, which are ubiquitous in nature and express one variable as a nonlinear function of another raised to a power [1,2]. A classic example in the domain of language is Zipf's law, which characterizes the frequency distribution of words in the form of a simple power law: $f \propto i^{-\alpha}$, where $f$ is the frequency of a word, $i$ its frequency rank, and the scaling exponent $\alpha \approx 1$.

A possible theoretical framework for understanding multiscale organization is provided by the criticality hypothesis. In complex systems, criticality refers to a system's functioning near a critical point, a 'sweet spot' between order (regularity) and disorder (chaos), where functioning is optimized and special formal properties emerge that provide empirical signatures of this particular regime. Key among these is the presence of long-range correlations, i.e. statistical temporal dependencies that decay slowly and span multiple orders of magnitude in time, often following power-law scaling [3]. In such a system, the current state of the system depends not only on recent history but on states of the distant past. Long-range temporal correlations thus reflect memory in the system, with past dynamics constraining current and future dynamics. This fits the structure of language, where texts exhibit long-range correlations driven by semantic information [4–7]. When telling a narrative, our utterances depend not only on what we just said (local correlations) but also on concepts we introduce before. In this instance, long-range correlations capture the fractal structure of discourse, in that semantic choices at any point are shaped by the accumulated weight of all previous choices, though whether their influence decays as a power law as opposed to exponentially is an open question.

Long-range correlations can be quantified through scaling exponents that characterize how fluctuations at different timescales relate to one another. Detrended Fluctuation Analysis (DFA) allows for the detection of scaling laws in non-stationary time series, removing local trends and revealing intrinsic correlations across multiple timescales that persist despite non-stationarity [8]. Specifically, if a time series exhibits scale-invariance, the magnitude of fluctuations grows as a power law (Equation 1)

$$F(n) \propto n^{\alpha} \qquad \textit{Equation } 1$$

where $F(n)$ is the fluctuation magnitude at timescale $n$. The value of the scaling exponent α provides a measure of the temporal structure in a signal. Values of $\alpha = 0.5$ correspond to uncorrelated, random fluctuations characteristic of white noise, indicating a lack of long-range correlations. Values above this indicate the presence of long-range correlations, where fluctuations at different time scales are systematically related. Higher values indicate increased self-similarity and persistence in the signal.

The presence of scale-free dynamics appears to be a hallmark of healthy physiological function across domains [9,10]. Healthy physiological signals, like heartbeat dynamics, can exhibit complex, long-range, power law correlations across multiple timescales, while disease and aging may be accompanied by disruptions of this multiscale organization [11]. Pathology has been shown to produce two opposite kinds of complexity loss: deviations towards white noise ($\alpha = 0.5$), which indicate a loss of long-range correlations and temporal structure as seen in the gait of Huntington's disease patients [12] and in heart rate variability measures in patients with atrial fibrillation [11,13], with shifts towards smoother, more persistent dynamics approaching brown noise ($\alpha = 1.5$) , as seen in the gait of patients with Parkinson's disease [14] as well as in congestive heart failure [11,15], which are indicative of excessive order or pathologic periodicity. Comparable alterations have been reported for brain activity in psychosis [16,17], including increased temporal persistence in MEG signals [18], and higher DFA scaling exponents in resting-state fMRI global signal of individuals with early psychosis [19]. If scale-free dynamics are a general feature of healthy functioning, they may also characterize language, instantiating overarching principles of computational optimality. In that case, temporal scaling in language could offer an accessible way of studying altered brain-functional dynamics.

Signatures consistent with scale-free organization have been documented across distinct levels of language and speech. At the acoustic level, power law scaling is observed in the distribution of energy releases during speech production across languages [20]; at the lexical level, Zipf's law as mentioned above is theorized to be a critical transition point in optimization of communication [21]. Zipf's and Heaps' laws, together with a stochastic model of text growth, have been described in written Chinese and English texts [22]. Such distributional properties, though, say little about how language unfolds over time. Natural language grammars produce a power-law decay in mutual information between symbols, in contrast to the exponential decay characteristic of simpler Markov processes [23], reflecting the presence of long-range structure and dependencies across scales. The temporal arrangement of linguistic units also shows long-range correlations directly. In texts, letter and word occurrences are long-range correlated and these correlations have been suggested to trickle down from the hierarchical organization of topics [4]. Coarse-grained texts, where letters are mapped onto a binary sequence, are anti-persistent at short distances and persistent at long distances, effects attributed to grammatical rules and semantics, respectively [6]. Further, long-range memory is found in the sequence of rare words and the return intervals between them across languages [7]. In one study, DFA has been applied to semantic similarity time series from conversation, where the cosine similarity between successive turns yielded exponents between white and pink noise [5]. Observations of such scaling signatures across multiple levels of language are consistent with critical-like organization in the domain of language, although a more complete theoretical account of language criticality remains to be developed.

In the context of psychopathology, language as a clinical and neuropathological signal has come to play an important role [24]. In psychosis, the disease process affects the lexical and the structural complexity of language (syntax) as well as the organization of meaning, at both lexical and grammatical levels, from early stages [25,26]. Recent work using current AI-based language models has studied this by vectorizing words in a high-dimensional semantic space, where the semantic similarity between words can be quantified by the cosine similarity between these vectors [27]. Across multiple studies and languages, there is now consistent evidence of a more compact ('shrunk') semantic space in psychosis, as seen initially through higher mean semantic similarity of words produced in connected speech [28–32], but more recently through a smaller hypervolume of the convex hull encompassing the embeddings of the words produced in a given text [33] and a more compressed semantic space [34–36]. Most NLP metrics applied in this domain, however, have been static ones targeting either local transitions or global metrics, missing out on a measure of temporal organization and being insensitive to dependencies that span long stretches of discourse.

Here, we apply DFA to semantic similarity time series derived from clinical transcripts to explore whether patients will show altered temporal scaling properties relative to controls. Our aim was a proof of concept in the linguistic domain, with a validation of an altered scaling exponent across independent samples and languages. We also asked whether scaling exponents related to symptom severity, for further clinical validation and because increased persistence in brain signals has been speculatively linked to the sustained nature of symptoms like delusions and hallucinations and the inability to break from these [18].

## 2. Methods

*Datasets*

Three datasets – here referred to as Zürich (Switzerland), Marburg (Germany), and Izmir (Türkiye) – were used in this analysis. Demographics for all samples are summarized in Table 1, along with differences between groups, calculated only for the analytical sample. The Zürich dataset [37] is comprised of recordings from 120 Swiss-German speakers: 40 patients with schizophrenia (SSD), 40 individuals with low schizotypy scores (LS) and 40 individuals with high schizotypy scores (HS). From these, 20 SSD, 29 LS, and 36 HS had at least one continuous transcribed speech segment of 101 or more words, required for stable DFA estimation, and were retained for analysis. Speech was elicited from these individuals using the discourse protocol (https://discourseinpsychosis.org/). To obtain transcripts, audio recordings underwent automatic speaker diarization and were aligned with textual transcripts using WhisperX [38]. Transcripts were then manually reviewed. Psychopathology was assessed using the Positive and Negative Syndrome Scale, PANSS [39] and formal thought disorder using the Thought, Language, and Communication scale, TLC [40]. All participants provided written informed consent, the ethics committee of Kantonale Ethikkommission Zürich gave ethical approval for this work (KEK-ZH 2020/01049), and the study adhered to the Declaration of Helsinki.

The Marburg dataset [33] is comprised of recordings from 129 German speakers: 43 patients diagnosed with schizophrenia and schizoaffective disorder (SSD), 43 patients diagnosed with major depressive disorder (MDD) (according to SCID-I, DSM-IV-TR), and 43 healthy controls (HC). From these, 33 HC, 34 individuals with MDD, and 39 individuals with SSD had at least one continuous speech segment of 101 or more words, required for stable DFA estimation, and were retained for analysis. In order to capture symptomatology, the Scale for the Assessment of Negative Symptoms, SANS [41], and the Scale for the Assessment of Positive Symptoms, SAPS [42], were used to examine clinical factors related to negative and positive symptoms respectively. Speech was elicited through a picture description/story telling task: the Thematic Apperception Test, TAT [43], using the methodology outlined for assessing the Thought and Language Index, TLI [44]. Transcriptions were conducted manually. Ethical permissions for the data collection and sharing of the data for the present study were obtained from the ethics committee of the Marburg University, School of Medicine (AZ 07:14).

The Izmir dataset is comprised of 148 native Turkish speakers: 57 patients with chronic schizophrenia (SZH), 17 patients with schizoaffective disorder (SZA) and 36 patients with first episode psychosis (FEP), as well as 38 healthy controls (HC). From these, 51 SZH, 16 SZA and 32 FEP patients, as well as 38 HC had at least one continuous speech segment of 101 or more words, required for stable DFA estimation, and were retained for analysis. Participants were recruited through the Psychotic Disorders Outpatient Unit of the Department of Psychiatry. Advertisements at the university campus and at the university hospital were used to include HCs who had no family or personal history of mental illness. The recruitment was approved by the Ethics Committee of Dokuz Eylul University. The protocol followed for the interviews is the Structured Clinical Interview for DSM-IV Axis I Disorders. In order to capture symptomatology, several symptom scales were used: Personal and Social Performance scale (PSP) [45], the Hamilton Depression Rating Scale (HDRS) [46], the Brief Negative Symptom Scale (BNSS) [47], as well as the SAPS. The dataset includes transcriptions of three different tasks: picture description, free speech, and retelling. The picture description task consists of describing three different images. Similarly, the free speech task includes conversational speech about participant's past, personal narrative, health narrative. Transcriptions were conducted manually.

*Analysis*

Participant speech was split into speech segments of continuous speech uninterrupted by interviewer utterances. All words as well as punctuation and capitalization were retained. Word-level semantic embeddings were obtained using Qwen3-Embedding-8B [48], a decoder only transformer (36 layers, 4096 dimensions). Each speech segment was embedded separately and passed in a single forward pass. Hidden states of the final layer were then extracted. Subword tokens were mapped back to whole words and each word's embedding was computed as the mean of the hidden states of its constituent sub-word tokens after the model's final normalization. Cosine similarity was then computed between adjacent word embeddings:

$$K(X,Y) = <X,Y> / \left(||X|| * ||Y||\right) \qquad \textit{Equation 2}$$

This is the standard measure to estimate similarity between vectors in a space [49], here adjacent words converted into embeddings. The resulting cosine similarity time series was analyzed using DFA to quantify long-range temporal correlations. This approach follows O'Nell & Finn [5], who applied DFA to semantic similarity time series to characterize temporal scaling in conversational speech. An overview of the pipeline is shown in Figure 1.

DFA consists of five steps. First, the time series is integrated by computing the cumulative sum of mean-subtracted values (Equation 3).

$$y(k) = \sum_{i=1}^{k}[x(i) - \bar{x}] \qquad \textit{Equation 3}$$

where $x(i)$ is the time series value at measurement $i$, $\bar{x}$ is the global mean of the series, and $k$ is the current position in the time series. Second, the integrated series $y(k)$ is divided into non-overlapping windows of equal length $n$. Third, within each window, a linear trend is fit by least squares regression and subtracted. This is the detrending step, by which local non-stationarities are removed, as these would artificially inflate the estimates of long-range correlations (Figure 1B, top). Then, the root mean square of the residuals $F(n)$ is computed in each window to quantify fluctuations at each scale. Finally, this procedure is repeated across 20 log-spaced window sizes ranging from a minimum of 4 to a maximum of one quarter of the segment length. Additionally, this procedure was applied to the reversed series to mitigate edge effects. The scaling exponent α is the slope of log $F(n)$ regressed on log $n$ (Equation 1; Figure 1B, middle). A value of $\alpha = 0.5$ indicates uncorrelated white noise with no long-range temporal structure. Values above 0.5 indicate the existence of long-range correlations, with increasing persistence as α approaches 1 (1/f noise), and values around 1.5 are characteristic of brown noise dynamics.

DFA often requires large amounts of data as long series are able to give more stable exponent estimates, with linear scaling over several orders of magnitude of window sizes. Still, DFA can be applied to much shorter series (down to about 64 points) [50] with the caveat that estimates of scaling become increasingly variable and unstable. Continuous speech segments of speech in clinical interviews are well-known to be generally limited quantitatively, in addition to being bounded by interview turns. In order to make use of clinical transcripts as well as ensure a stable and faithful estimation of α, we followed the recommendations of Kuznetsov & Rhea [51] for short or discontinued time series. Their work on DFA analysis of gait exposed similar constraints, as their clinical populations were not able to walk for extended periods of time. Through a simulation supported by empirical data, the authors demonstrated that group differences in DFA scaling coefficients (α) can be detected in repeated measurements of 100 to 200 points series, where multiple short series per participant are analyzed separately and their exponents are averaged, reducing the variance of the participant-level estimate. In line with that, we use 100 continuous data points (101 words) as a minimum of segment length. A scaling coefficient was then calculated for each qualifying speech segment, and these were averaged to yield a single participant-level DFA scaling coefficient (α).

Because a scaling exponent can be fitted to F(n) whether or not it follows a power law, we additionally compared the power-law model against linear and exponential alternatives for every speech segment, using BIC, out-of-sample prediction across scales, and calibration against simulated fractional Gaussian noise (Supplementary Methods S1).

DFA scaling exponents can also be biased by spikes, trends, and oscillations [52,53], so we quantified these in every similarity series (speech segment). Spikes were defined as points beyond 3 standard deviations from the median. Trends were quantified as the total linear drift across the series in units of its standard deviation and by Spearman correlation with time. Oscillations were tested with a periodogram peak test against a fitted 1/f background after linear detrending with a shuffled control as reference. Details are reported in Supplementary Methods S2. Because linear detrending (DFA-1) may be insufficient for some trends [53], we also repeated the analysis with quadratic detrending (DFA-2; Supplementary Methods S3).

*Shuffled-series control*

To test whether the scaling exponent genuinely depended on the temporal order of the similarity values, rather than on the distribution alone, each speech segment's cosine similarity series was randomly shuffled 2000 times, preserving the values but destroying the order (Figure 1A, bottom). DFA was applied to the shuffled series. For each speech segment, the mean α across shuffles was taken as its null exponent (Figure 1B, bottom). We also computed, for each segment, the proportion of its 2000 shuffled exponents falling below the observed α, and related this proportion to the observed α across segments within each dataset using Spearman's rank correlation. Participant-level shuffled exponents were obtained by averaging speech segment-level shuffled means in the same way as for observed exponents.

*Sensitivity analyses*

Along with DFA-2, we ran two additional sensitivity analyses: The first tested whether the choice of retaining or discarding disfluencies drove the results, hence we removed filler words and ran the same pipeline unchanged. This was motivated by evidence that filler use may itself differ in schizophrenia [54]. The second addressed the concern of insufficient sequence lengths and tested exploratorily whether concatenating speech across the entire transcript yielded the same results. This strategy provides a longer timeseries per participant but deliberately includes discontinuities in text due to interview questions or topic changes. In each case, the group models (Equations 4-6) were refitted (Supplementary Methods S3).

*Group comparisons*

The resulting participant-level DFA scaling exponents, averaged across tasks for stability as described above, were compared between diagnostic groups within each dataset using ordinary least squares linear regression models adjusted for number of words, age, and sex. Separate models were estimated for the Zürich, Izmir, and Marburg datasets (see Equations 4–6 for the corresponding model specifications). Models were implemented in python using the statsmodels [55] package (version 0.14.0). Across models, $\alpha_i$ is the DFA scaling exponent of participant $i$, $\beta_0$ is the expected α of the reference group, $n_{words}$ is the average number of words across a participants' speech segments, sex is coded as 0 = female and 1 = male and $\varepsilon_i$ is residual error. Diagnostic groups were entered as binary (0/1) variables and each coefficient is the adjusted difference in α between the group and reference:

$$\alpha_i = \beta_0 + \beta_{LS} \times LS_i + \beta_{HS} \times HS_i + \beta_{n_{words}} \times n_{words_i} + \beta_{age} \times age_i + \beta_{sex} \times sex_i + \varepsilon_i \qquad \textit{Equation } 4$$

where reference group is SSD (patients with schizophrenia spectrum disorders), LS = 1 for the low schizotypy control group and HS = 1 for the high schizotypy control group.

$$\alpha_i = \beta_0 + \beta_{FEP} \times FEP_i + \beta_{SZA} \times SZA_i + \beta_{SZH} \times SZH_i + \beta_{n_{words}} \times n_{words_i} + \beta_{age} \times age_i + \beta_{sex} \times sex_i + \varepsilon_i \qquad \textit{Equation } 5$$

where reference group is healthy controls (HC), FEP = 1 for first-episode psychosis, SZA = 1 for schizoaffective disorder, and SZH = 1 for chronic schizophrenia.

$$\alpha_i = \beta_0 + \beta_{MDD} \times MDD_i + \beta_{SSD} \times SSD_i + \beta_{nwords} \times nwords_i + \beta_{age} \times age_i + \beta_{sex} \times sex_i + \varepsilon_i \qquad \textit{Equation 6}$$

where reference group is healthy controls (HC), MDD = 1 for major depressive disorder, SSD = 1 for schizophrenia spectrum disorders.

*Symptom correlations*

To examine the relationship between DFA scaling exponents and symptomatology, partial Spearman correlations were computed between α and symptom scale scores across all participants, controlling for diagnosis. This allowed symptom correlations to be estimated across the full sample while accounting for the confounding effect of diagnostic group. To account for multiple comparisons, p-values were corrected using Benjamini-Hochberg false discovery rate (FDR).

## 3. Results

*Power-law scaling of the fluctuation function*

Because a scaling exponent can be fitted to F(n) whether or not it follows a power law, we first assessed whether the fluctuation functions were well described by a power law before comparing groups. Log-log fits of F(n) yielded high $R^2$ values (mean $R^2 = 0.9637$ (Zürich), 0.9727 (Marburg), 0.9734 (Izmir)). In a model comparison, a power law was preferred over linear and exponential alternatives by BIC in 98.96% (Zürich), 97.82% (Marburg), and 97.63% (Izmir) of speech segments, and was consistent with simulated power law surrogates in about 80% of speech segments (Supplementary Results S1; Supplementary Table 1). Given the short range of scales, this result supports that F(n) was well described by a power law across datasets, though not that the signals are strictly scale-free.

*Group comparisons*

Within each dataset, we fitted an adjusted linear regression model predicting DFA scaling exponents from diagnostic group, number of words, age and sex (Equations 4–6). Psychosis groups showed consistently higher scaling coefficients as compared to healthy controls across all datasets, indicating more persistent long-range temporal correlations in semantic similarity time series. In the Zürich dataset, the model was significant overall ($R^2 = 0.390$, adjusted $R^2 = 0.352$; $F(5,79) =$

10.11, $p < 0.001$). Using the SSD group as reference, both the low schizotypy group ($\beta_{LS}$ = -0.0295, SE = 0.009, p = 0.001) and the high schizotypy group ($\beta_{HS}$ = - 0.0263, SE = 0.008, p = 0.003) showed significantly lower scaling exponents. Among covariates, the number of words was negatively associated with the scaling exponent ($\beta_{nwords} = -0.0005$, $p < 0.001$), and age showed a positive association ($\beta_{age} = 0.0014$, p = 0.039). Sex showed a marginal effect, although not significant ($\beta_{sex} = 0.0113$, p = 0.092). These results are visualized in Figure 2A.

In the Izmir dataset, the model was significant overall ($R^2 = 0.210$, adjusted $R^2 = 0.173$; $F(6,130) = 5.744$, $p < 0.001$). Using healthy controls (HC) as the reference group, the SZH group showed a significant positive difference in DFA scaling exponents compared with HC ($\beta_{SZH}$ = 0.0303, SE = 0.011, p = 0.006), whereas the FEP ($\beta_{FEP} = 0.0229$, p = 0.078) and SZA ($\beta_{SZA}$ = 0.0206, p = 0.173) groups did not differ significantly from HC. Among covariates, the number of words was negatively associated with DFA scaling exponents ($\beta_{nwords} = -0.0003$, $p < 0.001$), while age ($\beta_{age} = 0.0005$, p = 0.275) and sex ($\beta_{sex} = -0.0041$, p = 0.640) showed no significant effects. These results are visualized in Figure 2B.

In the Marburg dataset, the corresponding model was significant overall ($R^2 = 0.223$, adjusted $R^2 = 0.184$; $F(5,100) = 5.741$, $p < 0.001$), indicating that the model explained a significant proportion of variance. Using HC as the reference category, the SSD group showed a significant positive difference in DFA scaling exponents compared with the reference group ($\beta_{SSD} = 0.0273$, SE = 0.011, p = 0.018), whereas the MDD group did not differ significantly from the reference ($\beta_{MDD} = 0.0087$, SE = 0.012, p = 0.451). Among covariates, the number of words was negatively associated with the scaling exponent ($\beta_{nwords} = -8.37 \times 10^{-5}$, p = 0.004), and age showed a significant positive association ($\beta_{age} = 0.0011$, p = 0.004). Sex showed no significant effects ($\beta_{sex}$ = $-0.0112$, p = 0.247). These results are visualized in Figure 2C.

Shuffled time series yielded exponents closer to 0.5 in all groups (Zürich: 0.5776 (SD: 0.005920) for LS, 0.5732 (SD: 0.005360) for HS, and 0.5763 (SD: 0.006644) for SSD, Izmir: 0.5756 (SD: 0.009160) for HC, 0.5791 (SD: 0.016732) for FEP, 0.5810 (SD: 0.016578) for SZA and 0.5811 (SD: 0.016388) for SZH, Marburg: 0.5633 (SD: 0.014633) for HC, 0.5671 (SD: 0.015385) for MDD and 0.5731 (SD: 0.014958) for SSD; Figure 2), as compared to the originally computed exponents. At the speech segment level, the proportion of shuffles below the observed $\alpha$ increased with $\alpha$: speech segments with an $\alpha$ near 0.5 were closer to their shuffled series whereas speech segments with a higher $\alpha$ exceeded (nearly) all shuffles (Spearman correlations: Zürich $\rho = 0.501$, $p < 0.001$, Marburg $\rho = 0.830$, $p < 0.001$, Izmir $\rho = 0.407$, $p < 0.001$).

The similarity series showed few features known to distort DFA [52,53] (Supplementary Table 2). Across datasets, spikes were sparse and isolated, linear drift was present but small, and oscillatory peaks were not more frequent than in the shuffled control series.

*Sensitivity analyses*

Group differences were largely robust to filler removal. All contrasts kept their direction and significance (Supplementary Table 3). DFA-2 also preserved all contrasts (Supplementary Table 4), though only with a minimum window of 8, chosen in order to prevent overfitting at small windows (Supplementary Methods S3). For the timeseries formed of concatenated segments across an entire transcript, group differences were retained in Marburg and Izmir but not Zürich (Supplementary Table 5).

*Symptom correlations*

*Zürich dataset*

Symptom correlations are displayed in Figure 3. Partial Spearman correlations controlling for diagnosis group revealed a significant positive association between DFA scaling exponents and PANSS negative symptoms ($\rho = 0.41$, $p_{FDR} < 0.001$), such that higher $\alpha$ was related to greater negative symptom severity. No significant associations were observed for any other subscale.

*Marburg dataset*

Symptom correlations are displayed in Figure 4. Partial Spearman correlations controlling for diagnosis group revealed a significant positive association between DFA scaling exponents and SAPS positive formal thought disorder ($\rho = 0.29$, $p_{FDR} = 0.031$, $n = 103$), such that higher $\alpha$ was related to greater thought disorder severity. No other SAPS or SANS measure survived FDR correction.

*Izmir dataset*

Partial Spearman correlations controlling for diagnosis group revealed no significant associations between DFA scaling exponents and symptomatology (Figure 5).

## 4. Discussion

The present study examined whether speech in psychosis exhibits alterations in the temporal scaling properties in their semantic similarity time series. Across all three datasets, patients with SSD had consistently higher DFA scaling exponents relative to healthy controls, indicating stronger long-range temporal correlations and thus, higher temporal persistence, in speech in psychosis. Several reliability analyses suggest that this result is largely robust to the choices and limitations of the method: exponents were well described by a power law over the available scales and were clearly separated from shuffled null series; spikes, trends and oscillations in the series were rare or small; and group differences were retained after removing filler words, with second-order detrending, and, in two of three datasets, in whole-transcript series. This is, to our knowledge, the first application of a scaling analysis to the temporal organization of semantic similarity in psychotic speech, suggesting that altered scaling properties previously found to mark pathologies in physiological domains [10,12,13,56], extend to language, and to psychosis in particular. Such alterations are consistent with, though they do not establish, a deviance in criticality, which in the case of psychosis has been independently evidenced for brain dynamics [18,19].

While this is a finding in the specific domain of the distribution of semantic distances between subsequent tokens, a question of future research is whether this finding is specific to this domain. If altered scaling reflects a systems-level property, we might see similar alterations reflected across a variety of system-relevant signal types (e.g., the time series of model-based token surprisal values) and other scaling signatures (e.g. Zipf's law of word frequency distributions). Boorom *et al.* [57] showed that parent-child dyads in autism spectrum disorder exhibited higher hierarchical temporal scaling, reflecting an increased clustering of speech and silences and decreased flexibility in the balance of vocal event timings compared to control dyads. Given findings of increased pressure of speech with positive symptoms of SSD and slower speech correlated with negative symptoms [58], individuals with SSD might also show such deviations in the scaling of vocal event timing. Interestingly, Luque *et al.* [20] show a power law in the inter-event distribution of energy releases in speech versus silences. As these authors speculate, this distribution concerns the intraphoneme range and may therefore indicate that mechanisms responsible for such criticality signatures are ultimately not cognitive but rooted in deeper, physical principles.

Another important question for future research is whether the alteration in language dynamics documented here could function as a readout of an altered neural dynamics under the same analytical framework. Lee *et al.* [16] reports disrupted power-law scaling in spontaneous neural activity and Fekete *et al.* [17] find that critical parameters shift away from optimal ranges in SSD. Using MEG, Alamian *et al.* [18] identified altered criticality signatures in SSD including increased temporal persistence in brain signals. The system thus shows increased memory, which is proposed to reduce efficiency of neural information processing [59] and is speculatively linked to the long and sustained quality of positive symptoms such as delusions and hallucinations [18]. Topal *et al*. applied DFA to resting state fMRI and found higher scaling exponents in individuals with early psychosis relative to healthy controls, accompanied by indices of weaker collective integration [19]. The same DFA metric as either applied to neural or linguistic dynamics thus shows a consistency in the direction of these effects. This supports the perspective that alterations in temporal scaling may cut across the brain and language domains, with language functioning as a readout of alterations of critical brain dynamics and its deviations in brain diseases. As first proposed in Ferrer-i-Cancho [21], reduced neural density in psychosis could be linked to deviations in the scaling properties of word frequency distributions using Zipf's law. More recently, evidence has emerged that automatically extractable language metrics relate to the functional and structural organization of the brain [60–62]. The present results extend this logic to the temporal domain, where scaling laws in cosine similarities may serve as readouts of underlying dynamical pathology.

Symptom correlations were scattered in the present study, covering both positive and negative symptom dimensions. The DFA scaling exponent may not index any particular symptom dimension but may instead capture a disruption in a broader systems-level principle of organization, which then manifests in different symptoms as downstream consequences. This may be consistent with Bleuler's distinction between 'fundamental processes', which he describes as primary disturbances characteristic of the underlying disorder, and 'accessory' symptoms, such as hallucinations and delusions, which are downstream effects of primary symptoms [63,64]. Rather than serving as a readout of its surface-level clinical manifestations, scaling exponents might thus reflect this deeper, primary disturbance.

This study has several limitations. DFA characterizes the correlational structure of a signal, i.e., how strongly values at different temporal windows are related. Signals can still share the same scaling exponent, though, and differ in their underlying dynamical organization [65]. A scaling exponent alone cannot fully characterize physiological complexity and thus, the interpretability of a single scaling exponent is limited. Interpretability could be much improved by contextualizing this exponent against other criticality signatures, capturing nonlinear magnitude correlations, analyzing multifractal spectra, as well as studying changes in relevant exponents over time and linking them to symptom changes [10]. However, speech quantities typically collected in most clinical studies limit the applicability of relevant methods. In the present study, in order to ensure a stable estimation of the scaling exponent, a minimum of 101 words (100 points) was set per segment, which led to a removal of a number of subjects across datasets. Here we took inspiration from Kuznetsov & Rhea [51] and obtained a less variable participant-level $\alpha$ by averaging the exponent of multiple observations per participant. Nevertheless, short series have important limitations: estimated exponents are more variable and subject to finite-size bias and reflect temporal persistence over only the available scales. Additionally, the range of available scales (~1 decade) cannot unambiguously distinguish a power law from exponential decay of correlations or short-memory processes [51]. Model comparisons were performed to distinguish these despite the short ranges. Although the evidence shows that a power law describes F(n) adequately over this range (Supplementary Results S1), it does not establish that this is the only adequate description. We therefore do not suggest that speech is a perfectly scale-free process, but rather that $\alpha$ is a meaningful descriptor of temporal structure and the clinical differences are reproducible.

It is also worth noting that, although the values of $\alpha$ match those in related literature [5], the observed scaling coefficient $\alpha$ values do not approach the 1/f scaling characteristic of canonical pink noise ($\alpha = 1$). Since both groups show exponents well below 1, and patients' exponents are closer to 1 than controls, the data are equally compatible with the reading that patients are nearer to a 1/f regime. Absolute exponents may depend on embedding model, window range and series length and thus we do not interpret these against a theoretical optimum. Word-level vector representations are an imperfect, noisy scalar summary of semantic content and the series per participant are short, which makes the DFA estimates more variable and unstable. Another important factor is box size configuration. The recommendation for maximum window size is usually one eighth or one tenth the length of the timeseries [52,53]. Here, we use one fourth to be able to cover a wider range of timescales in spite of our limited series length. Yet, since in this case there are substantially less observations at large windows, this may introduce additional variance at these longer timescales, thus lowering the exponents.

## 5. Conclusions

This study provides evidence of an alteration of the temporal organization of semantic similarity values in psychosis. We applied DFA to word-level cosine similarity time series across three independent datasets and found that patients with SSD exhibited significantly elevated scaling exponents relative to controls. These findings extend the framework of scale-free dynamics from neural activity to language, suggesting that altered scaling at the language level may serve as a readout of underlying changes in the dynamics of neural organization. They also suggest deeper explorations of how altered temporal organization speaks to language itself being a critical system, reflecting deviations from criticality signatures in brain diseases. Overall, this study presents evidence for a reproducible alteration of temporal scaling in contextual-semantic transitions in psychosis and raises the possibility that this may reflect altered dynamical organization of language, potentially related to critical-like neural dynamics.

*__Table 1:__ Demographics of participants and comparisons for each analytical sample*

| **Marburg** | **HC (mean ± SD)** | **MDD (mean ± SD)** | **SSD (mean ± SD)** | **MDD - HC (p-value)** | **MDD - SSD (p-value)** | **HC - SSD (p-value)** |
|---|---|---|---|---|---|---|
| number of original participants | 43 | 43 | 43 | | | |
| number of participants in analytical sample | 33 | 34 | 39 | | | |
| Age | 42.97 ± 12.92 | 40.85 ± 12.42 | 40.85 ± 12.11 | 0.4968 | 0.9981 | 0.4795 |
| Female (%) | 39.39% | 38.24% | 41.03% | 0.9225 | 0.8080 | 0.8881 |
| Education (years) | 15.03 ± 3.38 | 12.39 ± 2.28 | 11.92 ±1.98 | 0.0031** | 0.1589 | 0.0000*** |
| IQ | 117.23 ± 14.83 | 114.19 ± 12.77 | 112.49 ± 13.94 | 0.3912 | 0.6012 | 0.1884 |
| Age of onset (years) | - | 25.71 ± 12.64 | 17.63 ± 7.96 | - | 0.0037** | - |
| **Zürich** | **LS (mean ± SD)** | **HS (mean ± SD)** | **SSD (mean ± SD)** | **HS - LS (p-value)** | **HS - SSD (p-value)** | **LS - SSD (p-value)** |
| number of original participants | 40 | 40 | 40 | | | |
| number of participants in analytical sample | 29 | 36 | 20 | | | |
| Age | 25.03 ± 4.00 | 27.25 ± 4.64 | 27.25 ± 6.53 | 0.0429* | 1 | 0.1866 |
| Female (%) | 48.28% | 50% | 20% | 0.8902 | 0.0276* | 0.0436* |
| Education (years) | 11.97 ± 4.07 | 12.28 ± 4.03 | 9.15 ± 5.14 | 0.7583 | 0.0251* | 0.0484* |
| IQ | 97.93 ± 9.50 | 95.65 ± 6.81 | 94.2 ± 7.68 | 0.2920 | 0.4904 | 0.1404 |
| Age of onset (years) | - | - | 25.60 ± 7.04 | - | - | - |
| **Izmir** | **HC (mean ± SD)** | **FEP (mean ± SD)** | **SZA (mean ± SD)** | **SZH (mean ± SD)** | | |
| number of original participants | 38 | 36 | 17 | 57 | | |
| number of participants in analytical sample | 38 | 32 | 16 | 51 | | |
| Age | 34.92 ± 11.81 | 24.36 ± 8.85 | 37.94 ± 9.17 | 35.14 ± 10.81 | | |
| Female (%) | 34.21% | 50% | 50% | 37.25% | | |
| Education (years) | 13.79 ± 3.5 | 12.36 ± 3.84 | 12.56 ± 3.37 | 11.12 ± 3.11 | | |
| IQ | N/A | N/A | N/A | N/A | | |
| Age of onset (years) | - | 26.42 ± 9.67 | 24.31 ± 7.38 | 22.94 ± 6.69 | | |

| | **HC - FEP (p-value)** | **HC - SZA (p-value)** | **HC - SZH (p-value)** | **FEP - SZA (p-value)** | **FEP - SZH (p-value)** | **SZA - SZH (p-value)** |
|---|---|---|---|---|---|---|
| Age | 0.0000*** | 0.3193 | 0.9297 | 0.0000*** | 0.0000*** | 0.3164 |
| Female (%) | 0.1688 | 0.2771 | 0.7672 | 1.00 | 0.2363 | 0.3545 |
| Education (years) | 0.1086 | 0.2363 | 0.0004** | 0.8543 | 0.1256 | 0.1420 |
| IQ | N/A | N/A | N/A | N/A | N/A | N/A |
| Age of onset (years) | - | - | - | 0.4703 | 0.0755 | 0.2650 |
| | | | | | | |

Sex comparisons were performed using the $\chi^2$ test, while t-tests were used for comparing all other variables. All comparisons are done for the analytical samples. Significance levels are denoted as follows: * $p$ <.05, ** $p$ <.01, *** $p$ <.0001.

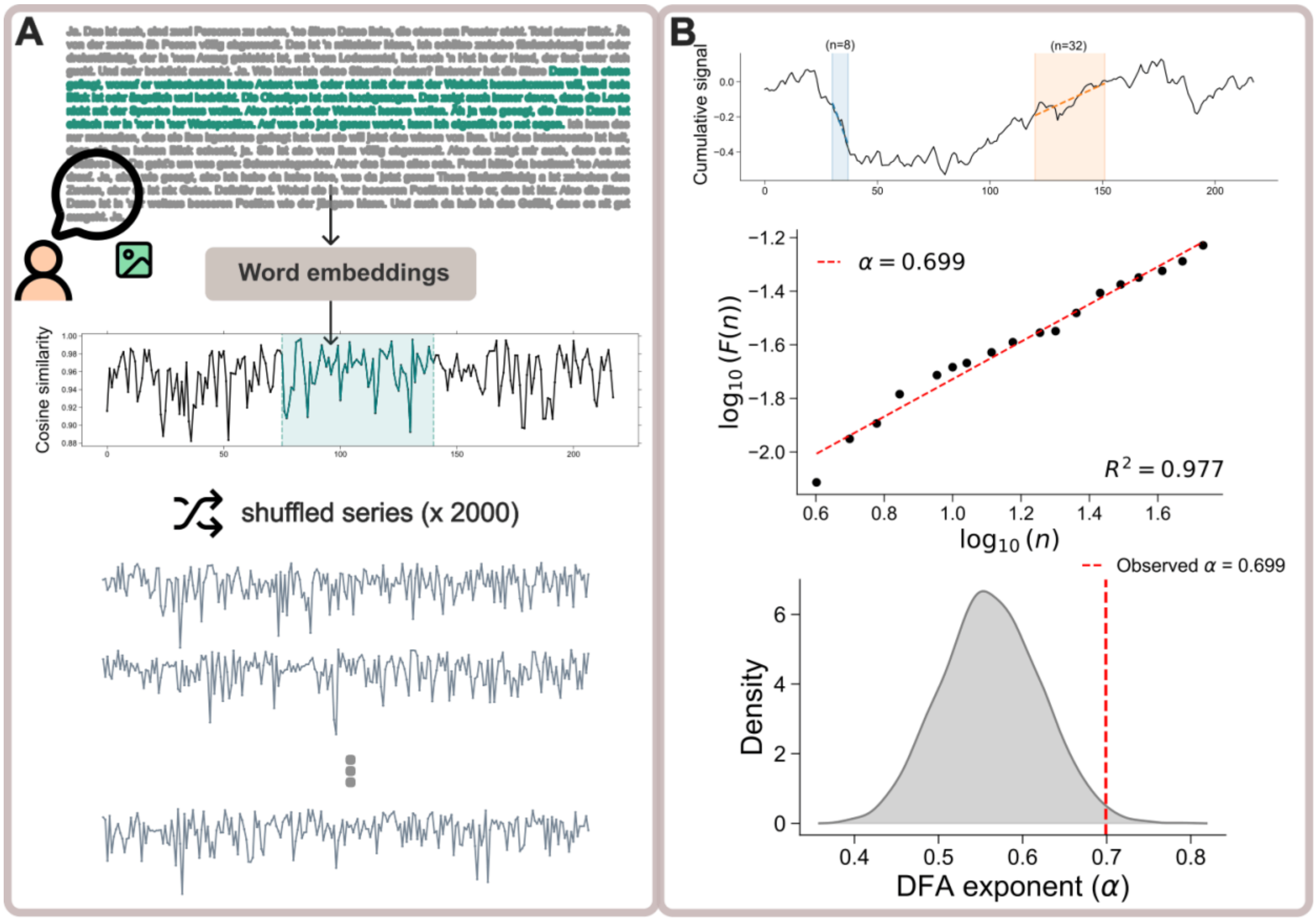

**Figure 1.** Analysis pipeline. (A) Schematic of signal constructing. Continuous stretches of participant speech, uninterrupted by interviewer utterances, are converted into contextualized word embeddings using Qwen3-Embedding-8B. Cosine similarity between consecutive word embeddings produces a time series representing successive semantic relationships throughout discourse. Detrended fluctuation analysis (DFA) is applied independently to each qualifying speech chunk to estimate its temporal scaling exponent \alpha. (B) Illustration of DFA and the shuffled-series control for an example speech chunk. The integrated, mean-centered cosine-similarity series is divided into windows of different sizes, within which local linear trends are removed. The resulting fluctuation function, F(n), is evaluated across window sizes n, and α is estimated from the slope of the log–log relationship. To assess the contribution of temporal ordering, the original cosine-similarity series is randomly shuffled 2000 times, preserving its values while disrupting their sequence. DFA exponents obtained from the shuffled series form a surrogate distribution against which the observed exponent is compared.

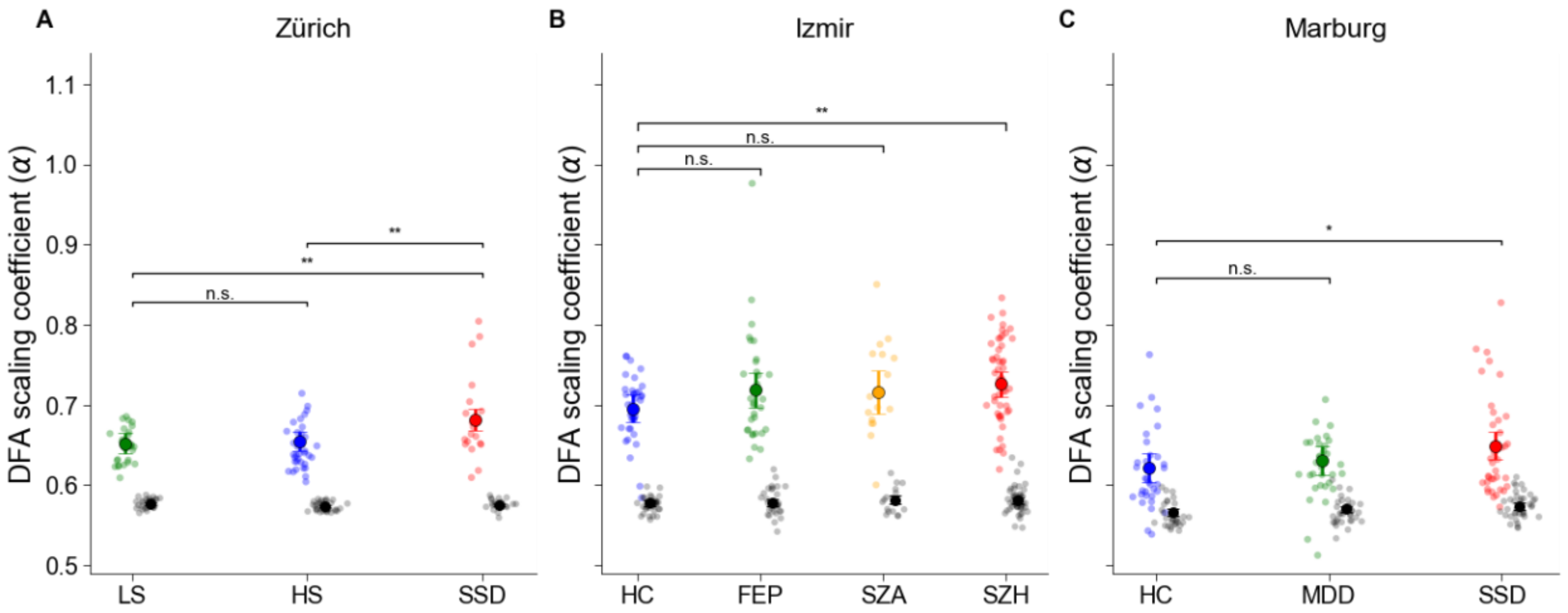

**Figure 2.** DFA scaling exponents (α) across diagnostic groups in the Zürich (A), Izmir (B), and Marburg (C) datasets. Scatter points represent individual participants: colored points represent observed exponents while grey dots show the corresponding exponents from shuffled versions of each participants' series. Error bars indicate model-estimated marginal means with 95% confidence intervals from adjusted linear regression models controlling for covariates. Brackets denote significant pairwise contrasts based on the fitted models (*$p < .05$ **$p < .01$, n.s. not significant).

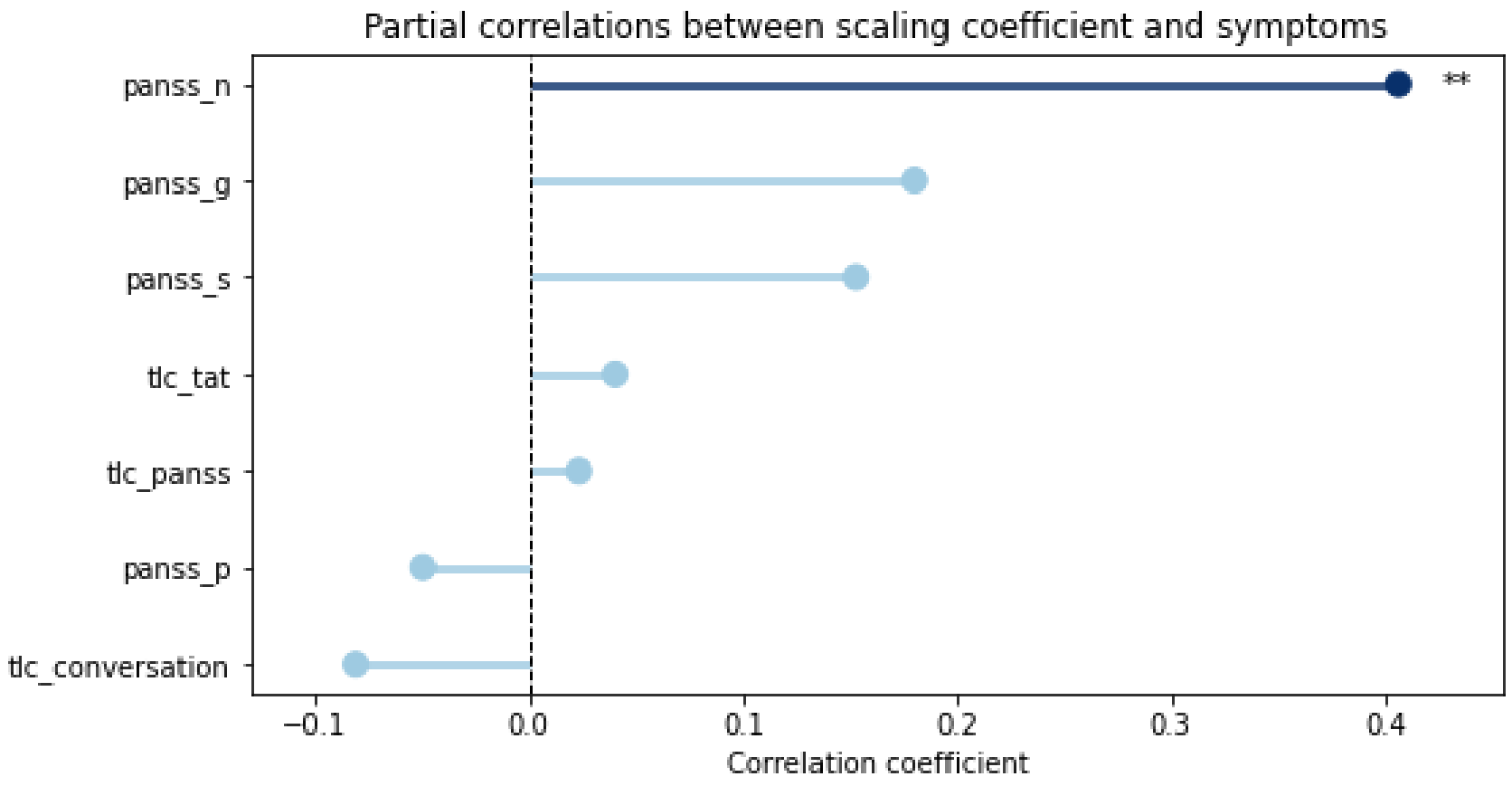

**Figure 3: Partial Spearman correlations between DFA scaling exponents (α) and symptom scores in the Zürich dataset, controlling for diagnosis.** Each point represents the correlation coefficient for a given symptom subscale, sorted by effect size. Dark blue indicates significance after FDR correction ($q < .05$), medium blue indicates uncorrected significance ($p < .05$), and light blue indicates non-significant associations. Dashed vertical line is at zero. *p < .05 (uncorrected); **q < .05 (FDR-corrected). Abbreviations: panss_p, panss_n, panss_g, are the positive, negative and general psychopathology subscales of the PANSS and panss_s is the summary subscore across all subscales.

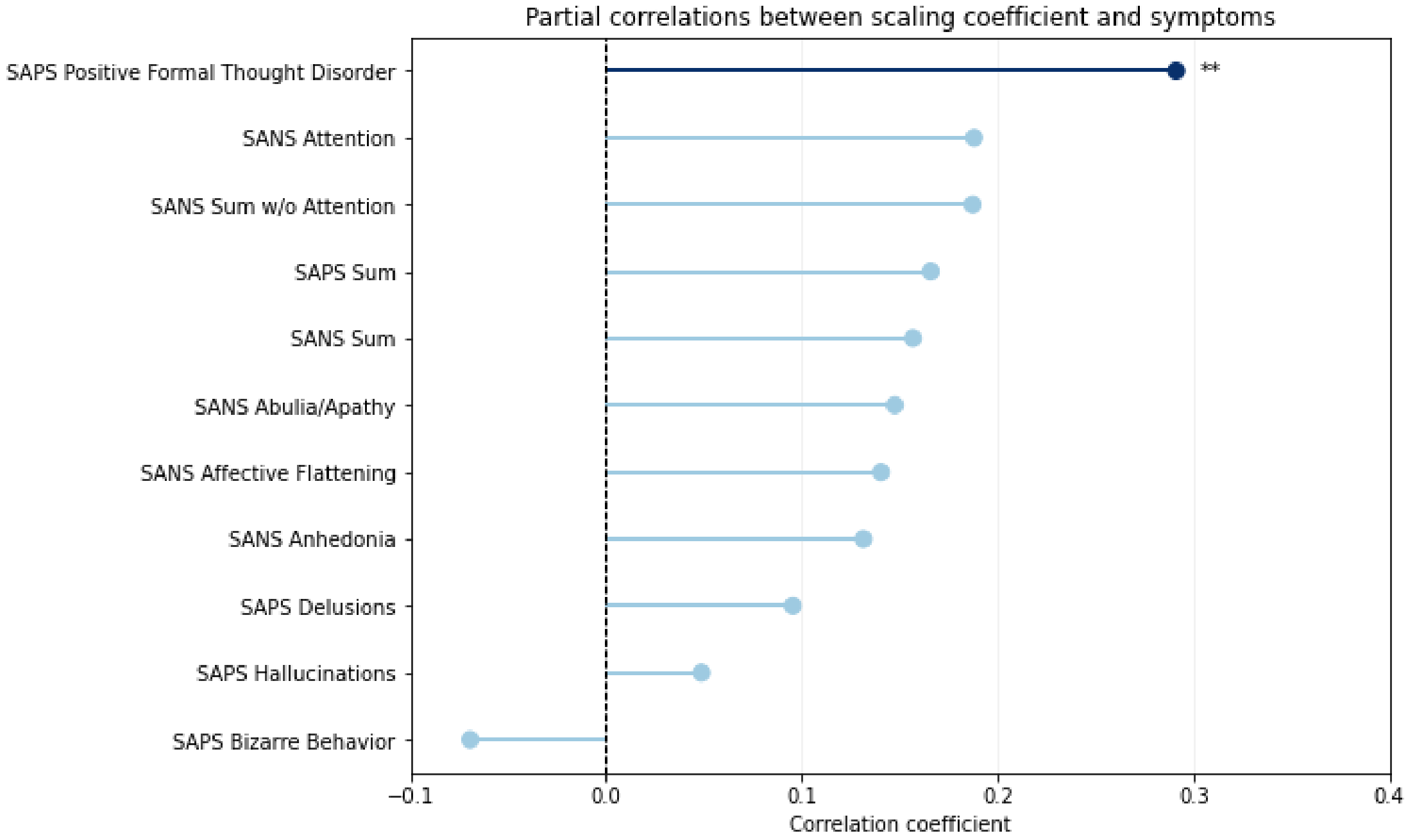

**Figure 4: Partial Spearman correlations between DFA scaling exponents (α) and symptom scores in the Marburg dataset, controlling for diagnosis.** Each point represents the correlation coefficient for a given symptom subscale, sorted by effect size. Dark blue indicates significance after FDR correction ($q < .05$), medium blue indicates uncorrected significance ($p < .05$), and light blue indicates non-significant associations. Dashed vertical line is at zero. *$p < .05$ (uncorrected); **$q < .05$ (FDR-corrected).

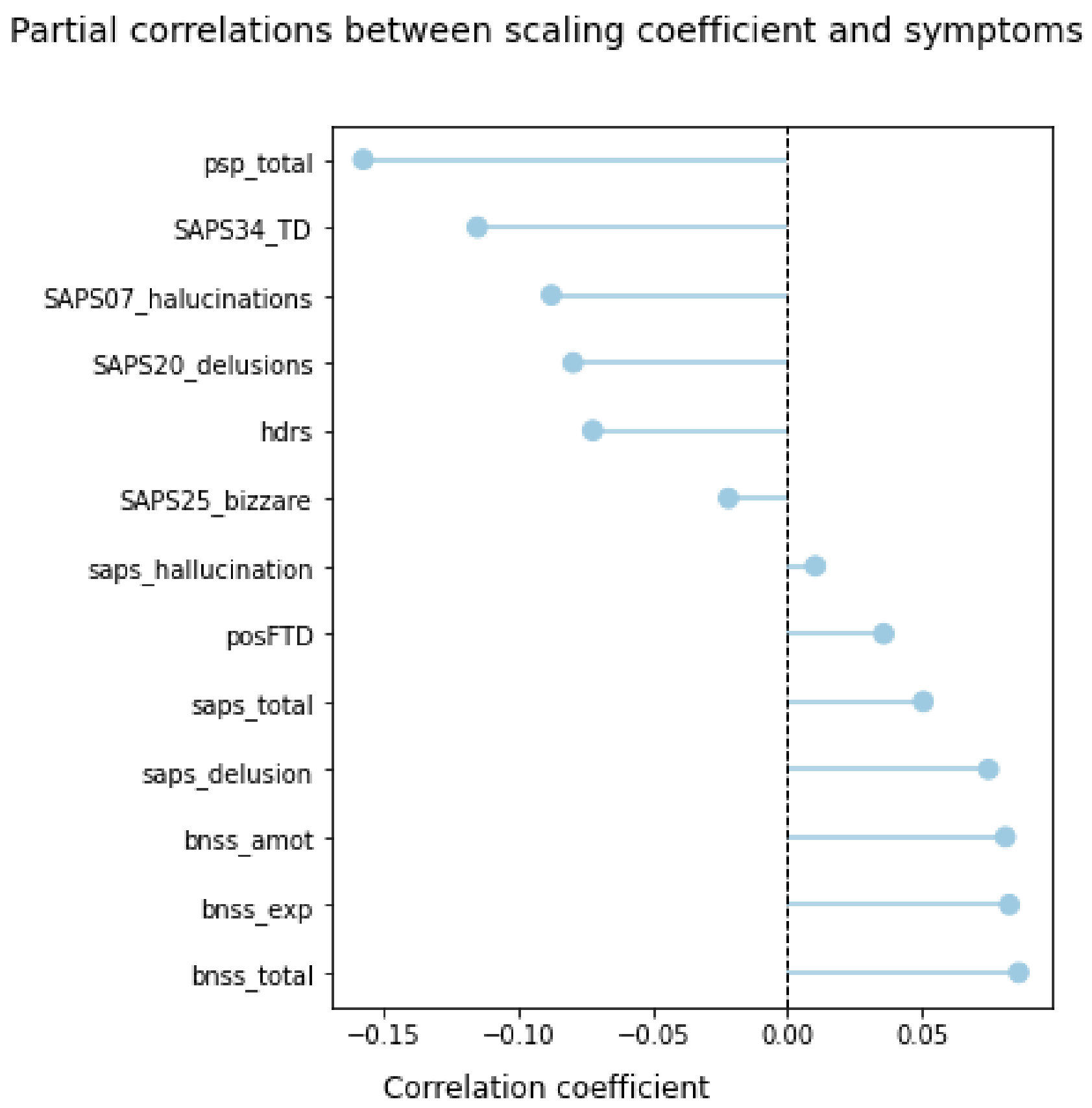

**Figure 5: Partial Spearman correlations between DFA scaling exponents (α) and symptom scores in the Izmir dataset, controlling for diagnosis.** Each point represents the correlation coefficient for a given symptom subscale, sorted by effect size. Dark blue indicates significance after FDR correction (q < .05**), medium blue indicates uncorrected significance (p < .05*), and light blue indicates non-significant associations. Dashed vertical line is at zero. *p < .05 (uncorrected); **q < .05 (FDR-corrected). Abbreviations: posFTD is positive formal thought disorder.

**Supplementary Methods**

*Supplementary Methods S1.* Assessing power-law scaling of F(n)

Analyses were run on each speech segment's cosine-similarity series separately (Zürich n = 671, Marburg n = 412, Izmir n = 379). DFA was used as described in the Methods. The log-log fit of F(n) was high (mean $R^2$ = 0.9637 (Zürich), 0.9727 (Marburg), 0.9734 (Izmir)). Because $R^2$ is high for many non-power law functions over short ranges, additional tests were done to assess scaling.

*Candidate models.* For each speech segment, three models were fitted to $log_{10}F(n)$ across all window sizes by least squares:

1. power law: $F(n) = A \cdot n^{\alpha}$, i.e. $log_{10}F \ = \ log_{10}A \ + \alpha \cdot log_{10}n$
2. linear: $\mathrm{F(n)} \ = \ \mathrm{A} \ + \ \mathrm{B} \cdot \mathrm{n}$, i.e. $log_{10}\mathrm{F} = \ log_{10}(\mathrm{A} \ + \ \mathrm{B} \cdot \mathrm{n}) \ with \ A, B \geq \ 0$
3. exponential: $F(n) = A \cdot e^{bn}$, i.e. $log_{10}F \ = \ log_{10}A + b \cdot log_{10}(e) \cdot \mathrm{n}$

All models were fitted to $log_{10}F(n)$ over the same window sizes.

*Model comparison.* Models were compared by the Bayesian Information Criterion, $BIC \ = \ N \cdot \ln(\mathrm{RSS/N}) + k \cdot \ln N$, where N is the number of window sizes, RSS the residual sum of squares and k = 2 for all models. The model with the lowest BIC was taken as preferred.

*Out-of-sample prediction*. Each model was fit to the smallest 70% of window sizes and used to predict $log_{10}F$ at the remaining larger window sizes. The model with the lowest prediction error (mean squared error in $log_{10}F$) was considered the best predictor.

*Calibration against fractional Gaussian noise.* For each segment, 200 fractional Gaussian noise series of an equal length were simulated with Hurst exponent H equal to the segment's estimated α, so that the surrogates are true power laws. All estimates fell within the range in which fGn is defined. The same DFA pipeline as described was applied to each surrogate. For each observed segment and each surrogate, we compare both the misfit of the log-log power law fit (i.e. proportion of surrogates with an RMSE of log-log residuals at least as large as observed) and the BIC advantage of the best alternative model over the power law (i.e. proportion of surrogates in which the best alternative's BIC advantage over the power law was at least as large as observed). We thus determined that a segment was labelled consistent with a power law if neither statistic exceeded the 95th percentile of the surrogate distribution.

*Supplementary Methods S2.* Spikes, trends and oscillations in the cosine similarity timeseries

DFA exponents can be affected by random spikes, trends, and oscillations in the signal [52,53]. We therefore characterize these features in the cosine similarity timeseries used for DFA.

*Spikes*. Each series was converted to robust z-scores, $z = (x - median)/(1.4826 \times MAD)$. Points with $|z| > 3$ were counted as spikes. For each series, we report the percentage of spike points, the largest $|z|$, the fraction of spikes that were isolated (no neighboring spikes), and skewness.

*Trends*. Linear drift was quantified as the total change of the least-squares lines across the series in units of the series SD (slope × length / SD). Monotonic trends were tested with Spearman's ρ between position and value.

*Oscillations.* For each series, the linear trend was removed and the periodogram was computed over frequencies up to 0.25 per word. A 1/f-type background was fitted to the periodogram by linear regression in log-log space and rescaled to the arithmetic mean. Under the null, periodogram-to-background ratios follow an exponential distribution.

*Supplementary Methods S3. Complementary analyses*

*Filler words.* Fillers and other disfluencies have been reported to differ between individuals with schizophrenia and controls [54], so in order to assess possible effects of filler words on the results, filler words were removed from the transcript text before embedding. In German and Swiss-German, this included words like äh, ähm, hm, mhm, and in Turkish words like ee, ıı, hmm. All embedding and DFA pipelines were identical to the main analysis. To test whether group differences still arise without fillers, we refitted the same group models (Equations 4, 5, and 6) for the similarity timeseries without fillers.

*DFA-order.* DFA-1 was used in the main analysis, but this removes constant offsets of the signal only and does not remove linear trends from the signal itself. We therefore repeated all analyses with second-order detrending (DFA-2), in which a quadratic polynomial is fitted within each window of the integrated series. But, higher-order detrending has a cost at short windows. A quadratic has three free parameters, so a 4-point window (as in our minimum window for the main analysis) leaves a single residual degree of freedom, and high-order polynomials are known to overfit the shortest blocks and in simulations, DFA-2 exponents were also more biased than DFA-1 exponents at short scales [66,67]. We therefore ran DFA-2 with two minimum window sizes: 4, as in the main analysis, and 8 (Supplementary Table 4). Group models (Equations 4, 5, and 6) were refitted on the resulting exponents (Supplementary Table 4).

*Concatenated series.* Due to the short nature of our speech segment timeseries, we also conducted a complementary analysis from concatenated whole transcripts. For each participant, all participant speech was retained and only interviewer utterances were removed. The text was embedded and the cosine-similarity series computed as in the main analysis and the same DFA pipeline was applied. This yields one longer time series per participant (mean 1784.5 words, range [754 - 4172] in Zürich; mean 1110.69 words, range [203 - 2494] in Marburg; mean 834.39 words, range [223 - 2377] in Izmir) but joins non-adjacent stretches as speech like separate picture descriptions or answers to interviewer questions, so they are not a continuous discourse. These discontinuities may cause edge effects, and the physical interpretation of correlation structure becomes more difficult [51]. Group models (Equations 4, 5, and 6) were refitted on the resulting exponents (Supplementary Table 5).

**Supplementary Results**

*Supplementary Results S1*

The power law was the preferred model by BIC in 664 of 671 speech segments used (98.96%) in Zürich, 403/412 (97.82%) of speech segments in Marburg, and 370/379 (97.63%) of the speech segments used in Izmir (Supplementary Table 1). In all remaining speech segments examined, the linear alternative was preferred, and the exponential model was never preferred.

When fitted to the smallest 70% of scales, the power law had the best prediction of the larger scales in 653/671 (97.3%) in Zürich, 394/412 (95.6%) in Marburg and 355/379 in Izmir (93.7%) of speech segments. The linear model was best in all other cases (Supplementary Table 1).

Relative to fGn surrogates, 80% (Zürich), 80% (Marburg), and 82% (Izmir) of segments were consistent with a power law on both criteria. Log-log misfit exceeded the surrogate expectation in 20%, 19%, and 17% of segments, whereas the alternative model did so in only 1%, 1% and 2% in respective datasets.

These results show that a power law describes F(n) at least as well as the alternatives tested. They do not show that the signals are strictly scale-free. Most window sizes here span less than one decade, and thus power-law and exponential decay of correlations are hard to distinguish at these lengths (see Discussion).

**Supplementary Table 1.** Power law assessment per dataset

| | Zürich | Marburg | Izmir |
|---|---|---|---|
| number of speech segments | 671 | 412 | 379 |
| mean $R^2$ (log-log) and [range] | 0.9637 [0.8626, 0.9953] | 0.9727 [0.8943, 0.9956] | 0.9734 [0.8720, 0.9961] |
| **preferred model by BIC** | | | |
| power law | 664 (98.96%) | 403 (97.82%) | 370 (97.63%) |
| linear | 7 (1.04%) | 9 (2.18%) | 9 (2.37%) |
| exponential | 0 (0%) | 0 (0%) | 0 (0%) |
| **Best out-of-sample prediction** | | | |
| Power law | 653 (97.3%) | 394 (95.6%) | 355 (93.7%) |
| Linear | 18 (2.7%) | 18 (4.4%) | 24 (6.3%) |
| Exponential | 0 (0%) | 0 (0%) | 0 (0%) |
| **Calibration against fGn surrogates ($p < .05$)** | | | |
| Log-log misfit larger than expected | 132 (20%) | 79 (19%) | 66 (17%) |
| Alternative-model advantage larger than expected | 5 (1%) | 5 (1%) | 6 (2%) |
| Consistent with a power law (neither criterion exceeded) | 534 (80%) | 330 (80%) | 309 (82%) |

*Supplementary Results S2*

*Spikes.* Few points deviated strongly from the median (Supplementary Table 2). The median percentage of spike points was 0% in Zürich, 0.9% in Marburg and 1.5% in Izmir, and 95% of series had at most 2.4%, 3.1% and 4.4%, respectively. Spikes were isolated (median isolated fraction 1.0 in all datasets) and moderate in size (median largest deviation 2.8–4.6 robust SDs; 95th percentile 4.6–6.1). Series were close to symmetric (median skewness −0.10 to 0.10).

*Trends.* Total linear drift was small: median absolute drift was 0.30, 0.19 and 0.26 SD, with 95th percentiles of 0.68, 0.62 and 0.80 SD (Supplementary Table 2). A monotonic trend was significant (Spearman $p < .05$) in 11%, 11% and 20% of series. Because the series are positively autocorrelated, this test rejects more often than its nominal 5% rate, so these percentages likely overstate the prevalence of true trends. We therefore rely mainly on the magnitude of the drift, which is small (median 0.19–0.30 SD).

*Oscillations.* A significant periodogram peak above the 1/f background was found in 8.5% (Zürich), 6.8% (Marburg) and 6.1% (Izmir) of series. These rates do not significantly exceed those obtained in shuffled copies of the same series (9.4%, 5.6% and 7.1%), and no consistent peak period emerged, so there was no evidence of oscillations beyond the false-positive rate of the test.

**Supplementary Table 2.** Characteristics of the cosine-similarity series.

| | Zürich | Marburg | Izmir |
|---|---|---|---|
| N series | 671 | 412 | 379 |
| Points per series, mean (range) | 129 (100-688) | 224 (100–833) | 171 (100-692) |
| **Spikes (\|z\| > 3 robust SD)** | | | |
| % of points, median (95th pct.) | 0 (2.4) | 0.9 (3.1) | 1.5 (4.4) |
| Largest \|z\|, median (95th pct.) | 2.8 (4.6) | 4.0 (5.2) | 4.6 (6.1) |
| Isolated fraction, median | 1.0 | 1.0 | 1.0 |
| Skewness, median | −0.10 | 0.10 | 0.08 |
| **Trends** | | | |
| Total drift (SD), median (95th pct.) of \|drift\| | 0.30 (0.68) | 0.19 (0.62) | 0.26 (0.80) |
| Median signed drift (SD) | 0.20 | 0.09 | 0.06 |
| Spearman trend p < .05 | 11% | 11% | 20% |
| **Oscillations** | | | |
| Periodogram peak p < .05 | 8.5% | 6.8% | 6.1% |
| Same test on shuffled copies | 9.4% | 5.6% | 7.1% |

*Supplementary Results S3.*

The results for all complementary analyses are summarized across Supplementary Table 3 - Supplementary Table 5, along with the original analysis for comparison.

*Filler words.* Group differences were largely unchanged (Supplementary Table 3). In all three datasets significant contrasts remained significant in the same direction.

*DFA order.* Under the original minimum window (4), contrasts in Marburg were retained as well as the SSD and HS contrast in Zürich. But the SSD and LS contrast in Zürich and the SZH and HC contrast in Izmir were both attenuated and were no longer significant, though the direction of the effect remained the same as for the original analysis. Using the larger minimum window (8), all group differences from the original analysis were preserved.

*Concatenated series.* When exponents were estimated from whole-transcript series as opposed to the average across segments as in the original analysis, group differences were retained in Marburg and Izmir. In Izmir, the FEP and HC contrast was not significant in the main analysis but became significant in this concatenated analysis to the same direction as SZH. In Zürich, group differences were not retained, with the direction of one contrast being in the opposite direction, though neither contrast was significant.

**Supplementary Table 3.** Comparison of analyses with and without filler words

| Dataset | **with fillers (main)** | | **without fillers** | |
|---|---|---|---|---|
| **Zürich (reference SSD)** | β (SE) | p | β (SE) | p |
| LS | - 0.0295 (0.009) | 0.001** | - 0.0333 (0.011) | 0.003** |
| HS | - 0.0263 (0.008) | 0.003** | - 0.0235 (0.010) | 0.027* |
| $n_{words}$ | - 0.0005 (0.00) | 0.000*** | - 0.0002 (0.00) | 0.084 |
| age | 0.0014 (0.001) | 0.039* | 0.0010 (0.001) | 0.241 |
| sex | 0.0113 (0.007) | 0.092 | 0.0092 (0.008) | 0.265 |
| **Izmir (reference HC)** | | | | |
| FEP | 0.0229 (0.013) | 0.078 | 0.0251 (0.018) | 0.156 |
| SZA | 0.0206 (0.015) | 0.173 | 0.0174 (0.021) | 0.397 |
| SZH | 0.0303 (0.011) | 0.006** | 0.0409 (0.015) | 0.006** |
| $n_{words}$ | - 0.0003 (0.00) | 0.000*** | -0.0002 (0.00) | 0.039* |
| age | 0.0005 (0.00) | 0.275 | 0.0006 (0.001) | 0.255 |
| sex | - 0.0041 (0.009) | 0.640 | - 0.0100 (0.012) | 0.405 |
| **Marburg (reference HC)** | | | | |
| MDD | 0.0087 (0.012) | 0.451 | 0.0183 (0.014) | 0.185 |
| SSD | 0.0273 (0.011) | 0.018* | 0.0287 (0.014) | 0.043* |
| $n_{words}$ | - 8.37 x $10^{-5}$ (0.00) | 0.004** | - 0.0002 (3.67 x $10^{-5}$) | 0.00*** |
| age | 0.0011 (0.00) | 0.004** | 0.0010 (0.00) | 0.047* |
| sex | - 0.0112 (0.010) | 0.247 | - 0.0210 (0.012) | 0.082 |

**Supplementary Table 4.** Comparison of analyses DFA-1 and DFA-2

| Dataset | **DFA-1 (main analysis)** | | **DFA-2 ($n_{min}$ = 4)** | | **DFA-2 ($n_{min}$ = 8)** | |
|---|---|---|---|---|---|---|
| **Zürich (reference SSD)** | β (SE) | p | β (SE) | p | β (SE) | p |
| LS | - 0.0295 (0.009) | 0.001** | - 0.0103 (0.010) | 0.306 | - 0.0320 (0.012) | 0.009** |
| HS | - 0.0263 (0.008) | 0.003** | - 0.0219 (0.010) | 0.024* | - 0.0324 (0.011) | 0.006** |
| $n_{words}$ | - 0.0005 (0.00) | 0.000*** | - 0.0010 (0.00) | 0.000*** | - 0.0007 (0.00) | 0.000*** |
| age | 0.0014 (0.001) | 0.039* | 0.0021 (0.001) | 0.006** | 0.0025 (0.001) | 0.007** |
| sex | 0.0113 (0.007) | 0.092 | 0.0091 (0.007) | 0.229 | 0.0109 (0.009) | 0.226 |
| **Izmir (reference HC)** | | | | | | |
| FEP | 0.0229 (0.013) | 0.078 | 0.0235 (0.014) | 0.084 | 0.0297 (0.016) | 0.058 |
| SZA | 0.0206 (0.015) | 0.173 | 0.0225 (0.016) | 0.156 | 0.0291 (0.018) | 0.111 |
| SZH | 0.0303 (0.011) | 0.006** | 0.0156 (0.011) | 0.168 | 0.0263 (0.013) | 0.045* |
| $n_{words}$ | - 0.0003 (0.00) | 0.000*** | - 0.0005 ($5.81 \times 10^{-5}$) | 0.000*** | - 0.0002 ($6.68 \times 10^{-5}$) | 0.000*** |
| age | 0.0005 (0.00) | 0.275 | 0.0004 (0.00) | 0.342 | 0.0005 (0.00) | 0.313 |
| sex | - 0.0041 (0.009) | 0.640 | 0.0065 (0.009) | 0.483 | - 0.0037 (0.011) | 0.724 |
| **Marburg (reference HC)** | | | | | | |
| MDD | 0.0087 (0.012) | 0.451 | 0.0268 (0.015) | 0.086 | 0.0183 (0.014) | 0.190 |
| SSD | 0.0273 (0.011) | 0.018* | 0.0458 (0.015) | 0.003** | 0.0378 (0.014) | 0.007** |
| $n_{words}$ | $- 8.37 \times 10^{-5}$ (0.00) | 0.004** | - 0.0002 ($3.84 \times 10^{-5}$) | 0.000*** | $- 8.86 \times 10^{-5}$ ($3.45 \times 10^{-5}$) | 0.012* |
| age | 0.0011 (0.00) | 0.004** | 0.0012 (0.001) | 0.017* | 0.0009 (0.000) | 0.043* |
| sex | - 0.0112 (0.010) | 0.247 | - 0.0208 (0.013) | 0.108 | - 0.0171 (0.012) | 0.141 |

**Supplementary Table 5.** Comparison of analyses of averaged speech segments and concatenated segments

| Dataset | **Averaged speech segments (main analysis)** | | **Concatenated segments** | |
|---|---|---|---|---|
| **Zürich (reference SSD)** | β (SE) | p | β (SE) | p |
| LS | - 0.0295 (0.009) | 0.001** | - 0.0075 (0.009) | 0.407 |
| HS | - 0.0263 (0.008) | 0.003** | 0.0032 (0.009) | 0.712 |
| $n_{words}$ | - 0.0005 (0.00) | 0.000*** | $- 2.107 \times 10^{-5}$ $(6.21 \times 10^{-6})$ | 0.001** |
| age | 0.0014 (0.001) | 0.039* | - 0.0004 (0.001) | 0.529 |
| sex | 0.0113 (0.007) | 0.092 | - 0.0099 (0.007) | 0.149 |
| **Izmir (reference HC)** | | | | |
| FEP | 0.0229 (0.013) | 0.078 | 0.0224 (0.010) | 0.024* |
| SZA | 0.0206 (0.015) | 0.173 | 0.0184 (0.011) | 0.111 |
| SZH | 0.0303 (0.011) | 0.006** | 0.0220 (0.008) | 0.009** |
| $n_{words}$ | - 0.0003 (0.00) | 0.000*** | $- 6.936 \times 10^{-5}$ $(8.42 \times 10^{-6})$ | 0.000*** |
| age | 0.0005 (0.00) | 0.275 | 0.0002 (0.00) | 0.515 |
| sex | - 0.0041 (0.009) | 0.640 | 0.0101 (0.007) | 0.130 |
| **Marburg (reference HC)** | | | | |
| MDD | 0.0087 (0.012) | 0.451 | 0.0118 (0.010) | 0.246 |
| SSD | 0.0273 (0.011) | 0.018* | 0.0226 (0.010) | 0.029* |
| $n_{words}$ | $- 8.37 \times 10^{-5}$ (0.00) | 0.004** | $- 1.476 \times 10^{-6}$ ($1.06 \times 10^{-5}$) | 0.889 |
| age | 0.0011 (0.00) | 0.004** | 0.0003 (0.00) | 0.339 |
| sex | - 0.0112 (0.010) | 0.247 | - 0.0079 (0.009) | 0.372 |